\documentclass[%
aps,%
prl,%
showpacs,%
twocolumn,%
preprintnumbers,
floatfix,%
superscriptaddress,%
nofootinbib%
]{revtex4}
\usepackage{graphicx,amsmath,amssymb}

\begin{document}


\preprint{ICRR-Report-562-2009-24}
\preprint{OUEPP-10-1}
\preprint{STUPP-10-206}

\title{
  A new idea to search for charged lepton flavor violation using a muonic atom
}

\author{Masafumi~Koike} %
\email{koike@krishna.th.phy.saitama-u.ac.jp} %
\affiliation{%
  Physics Department, Saitama University,
  255 Shimo-Okubo, Sakura-ku, Saitama, Saitama 338-8570, Japan
}
\author{Yoshitaka~Kuno}
\email{kuno@phys.sci.osaka-u.ac.jp} %
\affiliation{%
  Department of Physics, Osaka University,
  Toyonaka, Osaka 560-0043, Japan
}
\author{Joe~Sato} %
\email{joe@phy.saitama-u.ac.jp} %
\affiliation{%
  Physics Department, Saitama University,
  255 Shimo-Okubo, Sakura-ku, Saitama, Saitama 338-8570, Japan %
}
\author{Masato~Yamanaka} %
\email{yamanaka@icrr.u-tokyo.ac.jp} %
\affiliation{%
  Institute for Cosmic Ray Research,
  University of Tokyo, Kashiwa 277-8582, Japan %
}

\date{\today}

\begin{abstract}
We propose a new process of
$\mu^{-} \mathrm{e^{-}} \to \mathrm{e^{-}} \mathrm{e^{-}}$
in a muonic atom for a quest of charged lepton flavor
violation.
The Coulomb attraction from the nucleus in a heavy muonic atom leads to
significant enhancement in its rate, compared to $\mu^{+} \mathrm{e^{-}} \to
\mathrm{e^{+}}\mathrm{e^{-}}$.
The upper limit of the branching ratio is estimated to be of the orders of
$O(10^{-17}\,\textrm{--}\,10^{-18})$ for the photonic and the four Fermi
interactions from the present experimental constraints.
The search for this process could serve complementarily with the other relevant
processes to shed lights upon the nature of charged lepton flavor violation.
\end{abstract}


\maketitle

%
%
Charged lepton flavor violation (cLFV) is known to be one of the important rare
processes to search for new physics beyond the Standard Model (SM). Various
theoretical models predict sizable rates of cLFV processes, which are just below
the present experimental upper limits. The on-going and future experiments for
cLFV searches would reach sensitivities in the range of predictions by many
theoretical models. At this moment, the cLFV searches with muons present the
best limits owing to a large number of muons available for
measurements~\cite{Kuno:1999jp}. Typical cLFV processes with muons include
$\mu^{+} \rightarrow \mathrm{e^{+}} \gamma$, $\mu^{+} \rightarrow
\mathrm{e^{+}}\mathrm{e^{+}}\mathrm{e^{-}}$ and $\mu^{-}$ - $\mathrm{e^{-}}$
conversion in a muonic atom ($\mu^{-} \mathrm{N} \rightarrow \mathrm{e^{-}}
\mathrm{N}$). However, after the discovery of cLFV process in future, many other
different cLFV processes should be studied to shed lights upon understanding of
the nature of the cLFV interactions and develop insights into new physics
responsible for cLFV.

In this letter, we would like to propose a new cLFV reaction process of a bound
$\mu^{-}$ in a muonic atom, which is
\begin{equation}
  \mu^{-}\mathrm{e^{-}} \to \mathrm{e^{-}}\mathrm{e^{-}},
\label{eq:me2ee}
\end{equation}
where $\mu^{-}$ and
$\mathrm{e^{-}}$ in the initial state of Eq.(\ref{eq:me2ee}) are the muon and the atomic 1S
electron(s) bound in a Coulomb field of the nucleus in a muonic atom
respectively. 

This $\mu^{-} \mathrm{e^{-}} \rightarrow \mathrm{e^{-}} \mathrm{e^{-}}$ process
in a muonic atom has various significant advantages. First of all, this process
could have not only photonic dipole interaction but also four-Fermi contact
interaction, as in the processes of $\mu^{+}\rightarrow
\mathrm{e^{+}}\mathrm{e^{-}}\mathrm{e^{-}}$ and $\mu^{-} \mathrm{N} \rightarrow
\mathrm{e^{-}} \mathrm{N}$, whereas $\mu^{+} \rightarrow \mathrm{e^{+}} \gamma$
has only the former.  This would potentially allows us to investigate the full
structure of new physics beyond the SM. Secondly, this process has a two-body
final state, in which the two signal electrons are emitted almost back to back
and each of them has an energy of about a half the muon mass, $m_{\mu}/2$. This
would provide a cleaner experimental signature as well as a larger final-state
phase space than $\mu^{+} \rightarrow \mathrm{e^{+}} \mathrm{e^{+}}
\mathrm{e^{-}}$ decay.  Also, in comparison with the $\mu^{+} \rightarrow
\mathrm{e^{+}} \gamma$ search, the measurement of this process would be
relatively easier since no photon detection is involved.  Thirdly, one can
consider a similar reaction process with a muonium, such as $\mu^{+}
\mathrm{e^{-}} \rightarrow \mathrm{e^{+}}\mathrm{e^{-}}$. However, the rate of
this $\mu^{+} \mathrm{e^{-}} \rightarrow \mathrm{e^{+}}\mathrm{e^{-}}$ process
can not be large because of small overlap between the $\mu^{+}$ and
$\mathrm{e^{-}}$ wave functions.  However, in a muonic atom of atomic number
$Z$, we can increase the overlap between the $\mu^{-}$ and $\mathrm{e^{-}}$ wave
functions if an atom of large $Z$ is chosen. The enhancement occurs owing to the
Coulomb interaction from the nucleus which attracts the 1S state electron wave
function towards the $\mu^{-}$ and the nucleus.  The expected rate would
increase by a factor of $(Z-1)^{3}$. For example, the rate for a lead ($Z=82$)
is $5\times10^{5}$ times that of the $\mu^{+} \mathrm{e^{-}} \rightarrow
\mathrm{e^{+}}\mathrm{e^{-}}$ reaction. However, in a muonic atom, a nuclear
muon capture process occurs in addition to the normal Michel muon decay. But
since a lifetime of a muonic atom changes from 2.2 $\mu$s to $\sim 80 \,
\mathrm{ns}$ from a hydrogen to a lead, the reduction of the number of muons of
a factor of at most 20 can only be expected. Therefore, the net increase of the
branching ratio would become significant for a large $Z$. A potential
disadvantage is that the rate of the reaction process like this might not be
large enough compared to rare cLFV muon decays. Therefore, in this letter we
will evaluate the rate of this $\mu^{-} \mathrm{e^{-}} \rightarrow
\mathrm{e^{-}}\mathrm{e^{-}}$ process, and discuss its related issues.
%

\begin{figure}
  \includegraphics[width=87mm]{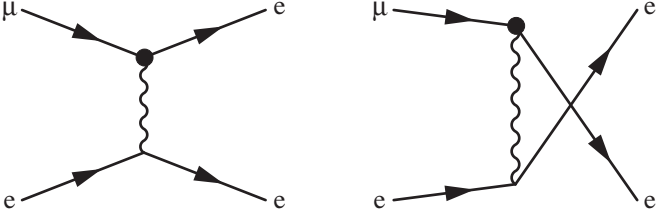}
  \caption{%
    The process $\mu^{-} \mathrm{e^{-}} \to \mathrm{e^{-}} \mathrm{e^{-}}$
    induced from the photonic interactions.
    The black dot indicates the effective interaction that is absent from the
    Standard Model.
  }
  \label{fig:mue-ee-diagrams}
\end{figure}

%
%
%
We describe the process of %
$\mu^{-}\mathrm{e^{-}} \to \mathrm{e^{-}}\mathrm{e^{-}}$ in a muonic atom of
Eq.~(\ref{eq:me2ee})
by an effective Lagrangian at the energy scale of the muon mass $m_{\mu}$.
Following Ref.~\cite{Kuno:1999jp}, we define
\begin{equation}
\begin{split}
  &
  \mathcal{L}_{\mu^{-}\mathrm{e^{-}} \to \mathrm{e^{-}}\mathrm{e}^{-}}
  =
  - \frac{4 G_{\textrm{F}}}{\sqrt{2}}
  \bigl[
    m_{\mu} A_{\textrm{R}} \,
    \overline{\mu_{\textrm{R}}} \sigma^{\mu\nu} e_{\textrm{L}} F_{\mu\nu}
    \\ & \hspace{0.5em}
    +
    m_{\mu} A_{\textrm{L}} \,
    \overline{\mu_{\textrm{L}}} \sigma^{\mu\nu} e_{\textrm{R}} F_{\mu\nu}
    \\ & \hspace{0.5em}
    + g_{1}
    ( \overline{\mu_{\textrm{R}}} e_{\textrm{L}} )
    ( \overline{  e_{\textrm{R}}} e_{\textrm{L}} )
    + g_{2}
    ( \overline{\mu_{\textrm{L}}} e_{\textrm{R}} )
    ( \overline{  e_{\textrm{L}}} e_{\textrm{R}} )
    \\ & \hspace{0.5em}
    + g_{3}
    ( \overline{\mu_{\textrm{R}}} \gamma^{\mu} e_{\textrm{R}} )
    ( \overline{  e_{\textrm{R}}} \gamma_{\mu} e_{\textrm{R}} )
    + g_{4}
    ( \overline{\mu_{\textrm{L}}} \gamma^{\mu} e_{\textrm{L}} )
    ( \overline{  e_{\textrm{L}}} \gamma_{\mu} e_{\textrm{L}} )
    \\ & \hspace{0.5em}
    + g_{5}
    ( \overline{\mu_{\textrm{R}}} \gamma^{\mu} e_{\textrm{R}} )
    ( \overline{  e_{\textrm{L}}} \gamma_{\mu} e_{\textrm{L}} )
    + g_{6}
    ( \overline{\mu_{\textrm{L}}} \gamma^{\mu} e_{\textrm{L}} )
    ( \overline{  e_{\textrm{R}}} \gamma_{\mu} e_{\textrm{R}} )
    \\ & \hspace{0.5em}
    + \textrm{(H.c.)}
  \bigr]
  \, ,
\end{split}
\label{eq:me-ee-Lagrangian}
\end{equation}
where $G_{\textrm{F}} = 1.166 \times 10^{-5} \, \mathrm{GeV^{-2}}$ is the Fermi
coupling constant, and $A_{\textrm{L, R}}$ and $g_{i}$'s ($i \in \{1, 2, \cdots,
6 \}$) are dimensionless coupling constants.
The first two terms in the brackets of Eq.~(\ref{eq:me-ee-Lagrangian}) are the
photonic interaction contributing to the process of Eq.~(\ref{eq:me2ee})
through the diagrams shown in Fig.~\ref{fig:mue-ee-diagrams}.
The remaining terms are for the direct four-Fermi contact interaction.

We estimate the branching ratio for the process of Eq.~(\ref{eq:me2ee}), which
takes place in a muonic atom with an atomic number $Z$.
The initial state has the muon and the electron in their 1S ground state of the
atomic orbits.
For simplicity, we ignore the three-momenta of the initial bound muon and
electron.
The final state has the two electrons, which, in the lowest order, can be
treated as monochromatic plane waves that propagate with opposite momentum
vectors.
Each of the two electrons in the final state takes energy of about $m_{\mu}/2$,
when the bound effects at the 1S state and the Coulomb interaction from the
nucleus can be neglected.
%

\begin{figure}
\begin{center}
  \includegraphics[width=87mm]{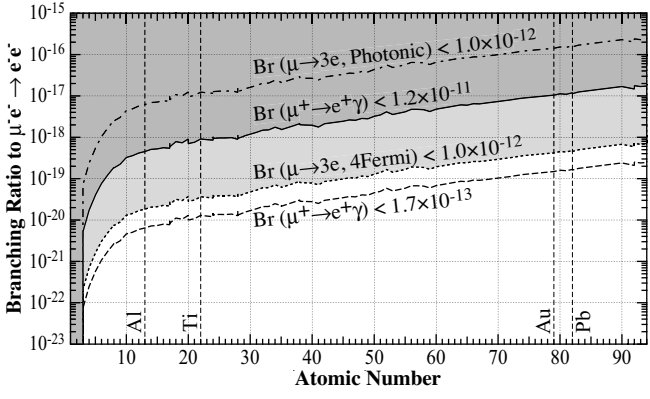}
  \caption{
    Limits on the branching ratio to the process of Eq.~(\ref{eq:me2ee})
    imposed by the limits on
    $\textrm{Br}(\mu^{+} \to \mathrm{e^{+}}\gamma)$ and on
    $\textrm{Br}(\mu^{+} \to \mathrm{\mathrm{e^{+}}\mathrm{e^{+}}\mathrm{e^{-}}})$.
    Models without photonic interactions are excluded in the
    light-shaded region.
    Similarly, models with photonic interactions are excluded in the dark-shaded
    region by the current experimental limits.
  }
\label{fig:br-limit}
\end{center}
\end{figure}

%
%
We begin with the first case where the four-Fermi interaction is dominant and
$A_{\textrm{R}}$ and $A_{\textrm{L}}$ are negligibly small, making no
contribution of the photonic interactions.
The remaining four-Fermi interaction allows the processes such as %
$\mu^{-}\mathrm{e^{-}} \to \mathrm{e^{-}}\mathrm{e^{-}}$ and %
$\mu^{+} \rightarrow \mathrm{e^{+}} \mathrm{e^{+}} \mathrm{e^{-}}$. %
The cross section of the process of Eq.~(\ref{eq:me2ee}) is calculated to be
\begin{equation}
\begin{split}
  &
  \sigma v_{\textrm{rel}}
  =
  \frac{1}{m_{\mu}^{2}}
  \frac{(G_{\textrm{F}}^{2} m_{\mu}^{2})^{2}}{16\pi} G
  \, ,
\end{split}
\label{eq:sv-4fermi}
\end{equation}
where
\begin{math}
  G \equiv G_{12} + 16G_{34} + 4G_{56} + 8G'_{14} + 8G'_{23} - 8G'_{56}
\end{math}
with
\begin{math}
  G_{ij} \equiv \lvert g_{i} \rvert^{2} + \lvert g_{j} \rvert^{2}
\end{math}
and
\begin{math}
  G'_{ij} \equiv \mathrm{Re} \, (g_{i}^{\ast}g_{j})
  \, .
\end{math}
The transition rate is then given by
\begin{equation}
\begin{split}
  &
  \Gamma(\mu^{-}\mathrm{e^{-}} \to \mathrm{e^{-}}\mathrm{e^{-}})
  =
  2 \sigma v_{\textrm{rel}}
  \bigl\lvert \psi_{\textrm{1S}}^{(\mathrm{e})}(0; Z - 1) \bigr\rvert^{2}
  \\ &
  =
  m_{\mu}  \frac{1}{8\pi}
  (Z - 1)^{3} \alpha^{3} (G_{\textrm{F}}^{2} m_{\mu}^{2})^{2}
  \Bigl( \frac{m_{\textrm{e}}}{m_{\mu}} \Bigr)^{3}
  G
  \, .
\end{split}
\label{eq:rate-4fermi}
\end{equation}
Here we took into account the facts that the 1S state can accommodate two
electrons, and that the nuclear charge is shielded by the negative muon.
We used the non-relativistic wave functions given by
\begin{equation}
  \psi_{\textrm{1S}}^{(\mathrm{e})}(r; Z - 1)
  =
  \frac{[(Z - 1) \alpha m_{\textrm{e}}]^{3/2}}{\sqrt{\pi}}
  \mathrm{e}^{- (Z - 1)\alpha m_{\textrm{e}} r}
  \, ,
\end{equation}
where %
$r$ is the radial coordinate, %
so that
\begin{math}
  \psi_{\textrm{1S}}^{(\mathrm{e})}(0; Z - 1)
  = [(Z - 1) \alpha m_{\textrm{e}}]^{3/2}/\sqrt{\pi} \, .
\end{math}
The rate of Eq.~(\ref{eq:rate-4fermi}) is enhanced for a larger atomic number,
$Z$, by a factor of $(Z - 1)^{3}$, giving a notable advantage for heavy nuclei.
This enhancement comes from the factor of $|\psi_{\textrm{1S}}^{(\mathrm{e})}(0;
Z - 1)|^{2}$, and the large positive charge of a heavy nucleus strongly attracts
the 1S wave functions of the leptons toward the nucleus position, rendering the
overlap of the two wave functions large, and enhances the transition of the
process of Eq.~(\ref{eq:me2ee}).
We normalize the rate of Eq.~(\ref{eq:rate-4fermi}) by the lifetime of a muonic
atom, $\tilde{\tau}_{\mu}$, to define the branching ratio of this process as
\begin{equation}
\begin{split}
  &
  \textrm{Br}(\mu^{-}\mathrm{e^{-}} \to \mathrm{e^{-}}\mathrm{e^{-}})
  \equiv
  \tilde{\tau}_{\mu}
  \Gamma(\mu^{-} \mathrm{e^{-}} \to \mathrm{e^{-}} \mathrm{e^{-}})
  \\ &
  =
  24\pi (Z - 1)^{3} \alpha^{3}
  \Bigl( \frac{m_{\textrm{e}}}{m_{\mu}} \Bigr)^{3}
  \frac{\tilde{\tau}_{\mu}}{\tau_{\mu}}
  G
  \, ,
\end{split}
\label{eq:br-4fermi}
\end{equation}
The value of $\tilde{\tau}_{\mu}$ ranges %
from $\tilde{\tau}_{\mu} = 2.19 \times 10^{-6} \, \mathrm{s}$ for
$\mathrm{^{1}H}$ %
to $\tilde{\tau}_{\mu} = (7 \textrm{ -- } 8) \times 10^{-8} \, \mathrm{s}$ for
$\mathrm{^{238}U}$ %
as listed in Ref.~\cite{Suzuki:1987jf}.
This is shorter than the lifetime of free muons, %
$\tau_{\mu} = 2.197 \times 10^{-6} \, \mathrm{s}$~\cite{Amsler:2008zzb},
which is equal to $192\pi^{3}/(G_{\textrm{F}}^{2} m_{\mu}^{5})$ at the lowest order.
The obtained branching ratio of Eq.~(\ref{eq:br-4fermi}) is to be compared with
that of $\mu^{+} \rightarrow \mathrm{e^{+}} \mathrm{e^{+}} \mathrm{e^{-}}$, the
process that arises from the same elementary process.
This branching ratio with no photonic interaction are given
by~\cite{Okada:1999zk}
\begin{equation}
\begin{split}
  &
  \textrm{Br}(\mu^{+} \to \mathrm{e^{+}} \mathrm{e^{+}}\mathrm{e^{-}})
  =
  \frac{1}{8} (G_{12} + 16G_{34} + 8G_{56})
  \, .
\end{split}
\label{eq:branch-m23e}
\end{equation}
The contribution from the interference among the four-Fermi interactions are
not present in Eq.(\ref{eq:branch-m23e}), whereas it is found to be present in Eq.~(\ref{eq:br-4fermi}) as the terms of
$G'_{ij}$'s.
The absence of the interferences is due to the large momenta of the final electrons.
The search for the process of Eq.~(\ref{eq:me2ee}) will thereby serve
complementarily with that for $\mu^{+} \rightarrow \mathrm{e^{+}} \mathrm{e^{+}}
\mathrm{e^{-}}$.
The ratios of the two branching ratios of
\begin{equation}
\begin{split}
  &
  \frac{\textrm{Br}(\mu^{-}\mathrm{e^{-}} \to \mathrm{e^{-}}\mathrm{e^{-}})}
       {\textrm{Br}(\mu^{+} \to \mathrm{e^{+}}\mathrm{e^{+}}\mathrm{e^{-}})}
  \lesssim
  192\pi
  (Z - 1)^{3} \alpha^{3}
  \Bigl( \frac{m_{\textrm{e}}}{m_{\mu}} \Bigr)^{3}
  \frac{\tilde{\tau}_{\mu}}{\tau_{\mu}}
  \, ,
\end{split}
\label{eq:ratio-of-ratios-m3e-4fermi}
\end{equation}
is obtained by assuming $G/(G_{12} + 16G_{34} + 8G_{56}) \sim O(1)$.
Hence, the upper limit on %
$\textrm{Br}(\mu^{-} \mathrm{e^{-}} \to \mathrm{e^{-}} \mathrm{e^{-}})$ %
is constrained by the existing limit of %
\begin{math}
  \textrm{Br}(\mu^{+} \to \mathrm{e^{+}}\mathrm{e^{+}}\mathrm{e^{-}})
  < B_{\textrm{max}}
\end{math}
as
\begin{equation}
\begin{split}
  &
  \textrm{Br}(\mu^{-}\mathrm{e^{-}} \to \mathrm{e^{-}}\mathrm{e^{-}})
  \\ & \hspace{4em}
  <
  192\pi
  (Z - 1)^{3} \alpha^{3}
  \Bigl( \frac{m_{\textrm{e}}}{m_{\mu}} \Bigr)^{3}
  \frac{\tilde{\tau}_{\mu}}{\tau_{\mu}}
  B_{\textrm{max}}
  \, .
\end{split}
\end{equation}
Figure~\ref{fig:br-limit} shows these upper limits as a function of an atomic
number $Z$ by the dotted curves. These upper limits are obtained by taking the
present experimental limit of $B_{\textrm{max}} = 1.0 \times 10^{-12}$ from the
SINDRUM experiment~\cite{mu-3e-expt}.
The light-shaded region in Fig.~\ref{fig:br-limit} is excluded by the current SINDRUM limit.
The reciprocal of the shown limit gives an estimation of the number of muons
that is required to detect events of the process of Eq.~(\ref{eq:me2ee}).
Let us take an example of the gold atom ($Z = 79$): %
our estimation requires a collection of %
$(4.21 \times 10^{-19})^{-1} = 2.38 \times 10^{18}$ muon events to improve the
current limit of $B_{\textrm{max}} = 1.0 \times 10^{-12}$.
Assuming the detection efficiency of $O(10\%)$, the required number of muons
amounts to a few times $10^{19}$.
Compared to this number is $O(10^{18}\,\textrm{--}\,10^{19})$ of muons, which is
the goal of the highly intense muon beams planned in the near future in search
for the cLFV~\cite{Mu2e,COMET,PRISM}.
We thereby find that the current limit could be reachable within the capability of
these muon sources.
Note that relativistic effects and the binding energy of the leptons may be
consequential for heavy nuclei, but are not included in our present estimation.

%
Let us now turn to the second case where the photonic interaction is present and
it dominates over the four-Fermi interactions.
In this case, the cross section, rate, and branching ratio of the
process of Eq.~(\ref{eq:me2ee}) are calculated to be
\begin{equation}
\begin{split}
  \sigma v_{\textrm{rel}}
  & =
  \frac{4 \alpha (G_{\textrm{F}} m_{\mu}^{2})^{2}}{m_{\textrm{e}}^{2}}
  \bigl(
    \lvert A_{\textrm{L}} \rvert^{2} +  \lvert A_{\textrm{R}} \rvert^{2}
  \bigr)
  \, ,
\end{split}
\label{eq:sv-photonic}
\end{equation}
\begin{equation}
\begin{split}
  &
  \Gamma(\mu^{-} \mathrm{e^{-}} \to \mathrm{e^{-}} \mathrm{e^{-}})
  =
  2 \sigma v_{\textrm{rel}}
  \bigl\lvert \psi_{\textrm{1S}}^{(\mathrm{e})}(0; Z - 1) \bigr\rvert^{2}
  \\ &
  =
  m_{\textrm{e}} \frac{8}{\pi} (Z - 1)^{3}
  \alpha^{4} (G_{\textrm{F}} m_{\mu}^{2})^{2}
  \bigl( \vert A_{\textrm{R}} \vert^{2} + \vert A_{\textrm{L}} \vert^{2} \bigr)
  \, ,
\end{split}
\label{eq:rate-photonic}
\end{equation}
and
\begin{equation}
\begin{split}
  &
  \textrm{Br}(\mu^{-} \mathrm{e^{-}} \to \mathrm{e^{-}} \mathrm{e^{-}})
  \\ &
  =
  1536 \pi^{2} (Z - 1)^{3} \alpha^{4}
  \bigl( \vert A_{\textrm{R}} \vert^{2} + \vert A_{\textrm{L}} \vert^{2} \bigr)
  \frac{m_{\textrm{e}}}{m_{\mu}}
  \frac{\tilde{\tau}_{\mu}}{\tau_{\mu}}
  \, ,
\end{split}
\label{eq:br-photonic}
\end{equation}
respectively.
On the other hand, the branching ratio of %
$\mu^{+} \to \mathrm{e^{+}} \mathrm{e^{+}} \mathrm{e^{-}}$ %
is given by~\cite{Okada:1999zk}
\begin{equation}
\begin{split}
  &
  \textrm{Br}(\mu^{+} \to \mathrm{e^{+}} \mathrm{e^{+}}\mathrm{e^{-}})
  \\ & \hspace{1em}
  =
  128 \pi\alpha
  \bigl( \lvert A_{\textrm{R}} \rvert^{2} + \lvert A_{\textrm{L}} \rvert^{2} \bigr)
  \Bigl[
    \log \Bigl( \frac{m_{\mu}}{m_{\textrm{e}}} \Bigr)^{2} - \frac{11}{4}
  \Bigr] \, .
\end{split}
\end{equation}
We then have
\begin{equation}
\begin{split}
  &
  \frac{\textrm{Br}(\mu^{-} \mathrm{e^{-}} \to \mathrm{e^{-}} \mathrm{e^{-}})}
       {\textrm{Br}(\mu^{+} \to \mathrm{e^{+}}\mathrm{e^{+}}\mathrm{e^{-}})}
  \\ & \hspace{1em}
  =
  12\pi (Z - 1)^{3}\alpha^{3}
  \frac{m_{\textrm{e}}}{m_{\mu}}
  \frac{\tilde{\tau}_{\mu}}{\tau_{\mu}}
  \Bigl[
    \log \Bigl( \frac{m_{\mu}}{m_{\textrm{e}}} \Bigr)^{2}
    - \frac{11}{4}
  \Bigr]^{-1}
  \, ,
\end{split}
\label{eq:ratio-of-ratios-m3e-Photonic}
\end{equation}
which we combine with the upper limit
\begin{math}
  \textrm{Br}(\mu^{+} \to \mathrm{e^{+}}\mathrm{e^{+}}\mathrm{e^{-}})
  < B_{\textrm{max}}
\end{math}
to yield
\begin{equation}
\begin{split}
  &
  \textrm{Br}(\mu^{-} \mathrm{e^{-}} \to \mathrm{e^{-}} \mathrm{e^{-}})
  \\
  & <
  12\pi (Z - 1)^{3}\alpha^{3}
  \frac{m_{\textrm{e}}}{m_{\mu}}
  \frac{\tilde{\tau}_{\mu}}{\tau_{\mu}}
  \Bigl[
    \log \Bigl( \frac{m_{\mu}}{m_{\textrm{e}}} \Bigr)^{2}
    - \frac{11}{4}
  \Bigr]^{-1}
  B_{\textrm{max}}
  \, .
\end{split}
\end{equation}
This upper limit is overlaid in Fig.~\ref{fig:br-limit} by a dash-dotted curve,
according to the aforementioned SINDRUM limit of $B_{\textrm{max}} = 1.0 \times
10^{-12}$.

%
The presence of the photonic interactions gives rise to another cLFV process %
$\mu^{+} \to \mathrm{e^{+}} \gamma$ as well, and search for this process also
put a limit to %
$\mathrm{Br}(\mu^{-} \mathrm{e^{-}} \to \mathrm{e^{-}} \mathrm{e^{-}})$.
The branching ratio to this process is given by
\begin{equation}
\begin{split}
  \textrm{Br}(\mu^{+} \to \mathrm{e^{+}} \gamma)
  & =
  \frac{\Gamma(\mu^{+} \to \mathrm{e^{+}} \gamma)}
       {\Gamma(\mu^{+} \to \mathrm{e^{+}}\nu_{\textrm{e}}\bar{\nu}_{\mu};
        \textrm{Free})}
  \\
  & =
  384\pi^{2}
  \bigl( \lvert A_{\textrm{R}} \rvert^{2} + \lvert A_{\textrm{L}} \rvert^{2}
  \bigr)
  \, ,
\end{split}
\label{eq:meg-br}
\end{equation}
which we compare with Eq.~(\ref{eq:br-photonic}) as
\begin{equation}
\begin{split}
  \frac{\textrm{Br}(\mu^{-} \mathrm{e^{-}} \to \mathrm{e^{-}} \mathrm{e^{-}})}
       {\textrm{Br}(\mu^{+} \to \textrm{e}^{+}\gamma)}
  &
  =
  4(Z - 1)^{3} \alpha^{4}
  \frac{m_{\textrm{e}}}{m_{\mu}}    
  \frac{\tilde{\tau}_{\mu}}{\tau_{\mu}}
  \, .
\end{split}
\label{eq:ratio-of-ratios-meg}
\end{equation}
Then the  limit on %
$\textrm{Br}(\mu^{-} \mathrm{e^{-}} \to \mathrm{e^{-}} \mathrm{e^{-}})$ %
is estimated from %
$\textrm{Br}(\mu^{+} \to \textrm{e}^{+}\gamma) < B_{\textrm{max}}$ as
\begin{equation}
\begin{split}
  \textrm{Br}(\mu^{-} \mathrm{e^{-}} \to \mathrm{e^{-}} \mathrm{e^{-}})
  &
  <
  \frac{\textrm{Br}(\mu^{-} \mathrm{e^{-}} \to \mathrm{e^{-}} \mathrm{e^{-}})}
       {\textrm{Br}(\mu^{+} \to \mathrm{e^{+}} \gamma)}
  B_{\textrm{max}}
  \\ &
  =
  4(Z - 1)^{3} \alpha^{4}
  \frac{m_{\textrm{e}}}{m_{\mu}}    
  \frac{\tilde{\tau}_{\mu}}{\tau_{\mu}}
  B_{\textrm{max}}
  \, .
\end{split}
\label{eq:18}
\end{equation}
A solid curve in Fig.~\ref{fig:br-limit} presents the upper limits given in
Eq.(\ref{eq:18}) with $B_{\textrm{max}}=1.2 \times 10^{-11}$, which is the
current upper limit from the MEGA experiment.
A dashed curve in Figure \ref{fig:br-limit} also shows the limits with the
$B_{\textrm{max}} = 1.7 \times 10^{-13}$, which is the goal value of the MEG
experiment~\cite{MEG}.
Even the current limit on the branching ratio to $\mu^{+} \to \mathrm{e^{+}}
\gamma$ overwhelms the limits on $\mathrm{Br}(\mu^{-} \mathrm{e^{-}} \to
\mathrm{e^{-}} \mathrm{e^{-}})$ when the photonic interaction is dominant.
Accordingly, the present excluded region is above the current MEGA limit.
Let us estimate, as we did earlier, the required number of muons to detect
events of the process of Eq.~(\ref{eq:me2ee}), taking an example of the gold
atom ($Z = 79$): %
an accumulation of $(1.044 \times 10^{-17})^{-1} = 9.58 \times 10^{16}$
muon events is necessary to surpass the sensitivity of MEGA, %
and %
$(1.480 \times 10^{-19})^{-1} = 6.76 \times 10^{18}$ to exceed that of the MEG
goal.
The required number of muons are estimated to be $10^{18}$ and a few of
$10^{19}$, respectively, assuming the $O(10\%)$ of the detection efficiency
again.
These are not capable now but would be possible in future with the planned
highly intense muon sources.

The expected magnitudes of the branching ratio of the process of
Eq.~(\ref{eq:me2ee}), which is driven by both the four-Fermi interaction and the
photonic interaction, are found to be not significantly large even with the
enhancement of $(Z-1)^{3}$.  Thus, this would not be the first process to
cultivate the discovery frontier of the cLFV searches. However, thanks to the
enhancement of $(Z-1)^{3}$, this process can be accessible in future by
next-generation high-intensity muon beams to produce muons of an order of
$O(10^{18} \,\textrm{--} 10^{19})$ per year.  On the other hand, the search for
$\mu^{+}\mathrm{e^{-}} \to \mathrm{e^{+}}\mathrm{e^{-}}$ would not be possible
with these intensities. Even higher beam intensities of $O(10^{21})$ muons per
year are envisioned in the future accelerator projects such as muon colliders,
neutrino factories, and Higgs factories~\cite{Geer:2010zz}.

A number of critical issues should be carefully studied to consider experimental
feasibility of this search.
First, one should take account of potential modification of energy spectra of
each of the emitted signal electrons due to the bound effects in a muonic atom.
It has been known that the electron spectrum from bound muon decays in a muonic
atom is strongly deformed for a large atomic number $Z$~\cite{Watanabe:1987su}.
For the process of Eq.~(\ref{eq:me2ee}), however, it is speculated that the sum
of the two signal electrons are still fixed to the initial energy of $m_{\mu} +
m_{\textrm{e}} - (\textrm{binding energies})$ due to the small kinetic
energy of the recoiled nucleus.
Secondly, the wave functions of the initial and final leptons should be improved
to take account of the nuclear charge distribution and relativistic effects.
A binding energy of the initial muon need to be also taken into account.
Thirdly, the interactions of the initial leptons need detailed treatments.
Various different models may predict the dependence of the rate on the initial
spin state. And the interaction between the two 1S electrons in the atomic
orbital of the muonic atom can be looked at.
Possible contributions from 2S, 3S, $\cdots$ states should also be considered.
Finally, an important experimental issue is the estimation of backgrounds, including the physical SM backgrounds from
\begin{math}
  \mu^{-}e^{-} \rightarrow \mathrm{e^{-}}\mathrm{e^{-}}\nu\bar{\nu}
\end{math}
decay, and accidental backgrounds that are known to be detector-dependent.
We will discuss these issues in our future works.

In summary, the new cLFV process $\mu^{-}\mathrm{e^{-}} \to
\mathrm{e^{-}}\mathrm{e^{-}}$ in a muonic atom is proposed. This process has the
rate enhancement of $(Z - 1)^{3}$ over the $\mu^{+}\mathrm{e^{-}} \to
\mathrm{e^{+}}\mathrm{e^{-}}$ owing to the Coulomb interaction from the nucleus
in a muonic atom.
This process has a final state of two electrons, which would be experimentally
very clear signature.
The upper limits of the branching ratio of the orders of $O(10^{-17}
\,\textrm{--} 10^{-18})$ are estimated separately for the photonic and the four
Fermi interactions from the other cLFV experimental results.
Once this process is observed, CP violation might be studied by comparing this process with $\mu^{+} \to \mathrm{e^{+}}\mathrm{e^{+}}\mathrm{e^{-}}$.

Acknowledgments.
The work of Y.~K. was supported in part by the Ministry of Education, Culture,
Sports, Science, and Technology, Government of Japan, Grant-in-Aid for
Scientific Research (A) (No.~20244029).
The work of J.~S. was supported in part by the Grant-in-Aid for the Ministry of
Education, Culture, Sports, Science, and Technology, Government of Japan
(Nos.~20025001, 20039001, and 20540251).
The work of M.~Y. was supported in part by the Ministry of Education, Culture,
Sports, Science, and Technology, Government of Japan, Grant-in-Aid for JSPS
Fellows (No.~20007555).

\end{document}